# Correlation between resistance fluctuations and temperature dependence of conductivity in graphene

VIERA SKÁKALOVÁ[1]*, ALAN B. KAISER[2], JAI SEUNG YOO[1], DIRK OBERGFELL[1] AND SIEGMAR ROTH[1]

[1]Max Planck Institute for Solid State Research, Heisenbergstrasse 1, 70569 Stuttgart, Germany
[2]MacDiarmid Institute for Advanced Materials and Nanotechnology, SCPS, Victoria University of Wellington, P O Box 600, Wellington, New Zealand

*e-mail: v.skakalova@fkf.mpg.de

**The weak temperature dependence of the resistance $R(T)$ of monolayer graphene[1-3] indicates an extraordinarily high intrinsic mobility of the charge carriers. Important complications are the presence of mobile scattering centres that strongly modify charge transport, and the presence of strong mesoscopic conductance fluctuations that, in graphene, persist to relatively high temperatures[4,5]. In this Letter, we investigate the surprisingly varied changes in resistance that we find in graphene flakes as temperature is lowered below 70 K. We propose that these changes in $R(T)$ arise from the temperature dependence of the scattered electron wave interference that causes the resistance fluctuations. Using the field effect transistor configuration, we verify this explanation in detail from measurements of $R(T)$ by tuning to different gate voltages corresponding to particular features of the resistance fluctuations. We propose simple expressions that model $R(T)$ at both low and high charge carrier densities.**

Recently, several papers have presented systematic analyses of $R(T)$ data in graphene[1-3]. Morozov et al.[1] ascribed the strong increase of resistivity above a temperature of 200 K to scattering by flexural phonons[6] localized in ripples in the graphene sheet. Chen et al.[2] fitted their $R(T)$ data to the usual linear term due to scattering by acoustic phonons and Bose-Einstein functions for scattering from specific high-energy phonons. They concluded that scattering by a pair of interfacial phonons in the $SiO_2$ substrate[7] was the most likely origin of the increase of $R(T)$ at high temperatures, but also noted that their resistivity increase at high $T$ was consistent with scattering by phonons of energy 104 meV. To eliminate the role of the substrate in limiting conductivity, Bolotin et al.[3] investigated a suspended graphene flake, achieving near-ballistic transport with mobility ~120,000 $cm^2$/Vs at 240 K. The resistivity increased linearly from 50 K to 240 K suggesting that longitudinal acoustic phonons were the main scatterers.

For low charge carrier densities near the charge neutrality point (NP), the resistivity is less well understood, generally increasing as temperature decreases but sometimes showing a decrease below 150 K[1,8]. Cho and Fuhrer[9] and Chen et al.[10] concluded that the resistance of their graphene samples near the neutrality point (NP) was governed not by the physics of the Dirac singularity but by carrier-density inhomogeneities induced by the potential of charged impurities ("puddles" of electrons and holes as imaged by Martin et al.[11]).

We have measured the resistance of monolayer graphene samples as a function of temperature and gate voltage (see Methods section for details). Figure 1 shows a typical resistance at $T$ = 4.2 K of a portion of graphene of length 1200 nm and width 1450 nm. The resistance as a function of gate voltage $V_G$ in Figure 1a shows the usual maximum in resistance (at the NP) with charge carrier density increasing for $V_G$ on either side of the maximum. The strong irregular fluctuations that decorate $R(V_G)$ at low temperature are an example of universal conduction fluctuations[12-18] which are observed in, for example, Si inversion layers, typically below helium temperatures[14,15]. In our case, these mesoscopic resistance fluctuations (MRFs) originate from the interference of electrons scattered from a particular distribution of scattering centres as the charge carrier density is changed by varying $V_G$. The interference pattern is reproducible on cycling $V_G$ at low temperature where the scattering centres are immobile.

At higher temperatures, some scattering centres may become mobile (as has been observed for adsorbed atoms on graphene sheets[19]) and rearrange along the sample. The interference pattern would then be changed when the sample is cooled down to low temperatures again. We infer that this happens during the $T$-cycle in Fig. 1b where the temperature was cycled up to 245 K and back to 4.2 K at constant $V_\Delta = V_G - V_{NP} = -0.6$ V, since $R(T)$ is different on the cooling part of the cycle. This is confirmed by the re-measurement of the MRF pattern at 4.2 K after the cycle (Figure 1a), which shows that the interference pattern is significantly changed. Zooming close to the maximum in the gate voltage dependence (inset of Fig. 1a), the value of the resistance at $V_\Delta = -0.6$ V on the black curve corresponds to the value of the resistance at 4.2 K in the $R(T)$ curve at the *beginning* of the $T$-cycle (black curve in Fig. 1b), and the value of the resistance at $V_\Delta = -0.6$ V on the grey curve matches the value of the resistance at 4.2 K in the $R(T)$ curve at the *end* of the $T$-cycle (grey curve in Fig. 1b). Resistance fluctuations are not seen directly in the resistance as a function of temperature at constant $V_G$, only the monotonic decay of each fluctuation feature.

We conclude that the dramatic initial drop of resistance as temperature is increased from 4.2 K for the heating curve represents the decrease of the MRF maximum as the phase coherence length decreases with rising temperature[13], leading to the reduction of the interference effects of the electron waves and the eventual disappearance of the MRF. To confirm this scenario, we made a systematic study of the correlation between the features in the resistance fluctuations and the temperature dependence of the resistance:

First, we measured $R(V_G)$ at various constant temperatures and calculated the root-mean-square of the resistance fluctuations $\Delta R_{rms}(T)$ by subtracting the value of $R(V_G)$ at temperature $T$ from that at a temperature of 51 K where the fluctuations are small. $\Delta R_{rms}(T)$ was found to follow closely an exponential decrease with temperature.

Second, we set constant values of the gate voltage corresponding either to a destructive electron waves interference (a MRF peak) or a constructive interference (a MRF valley) in $R(V_G)$ at 4.2 K and then measured $R(T)$ from 4.2 - 250 K. Values of $V_G$ chosen included low charge density near the NP (Figure 2), and high electron and hole densities (Figure 3). For all cases of $V_G$ set at a MRF peak, there is a sharp decrease in $R(T)$ as $T$ increases from 4.2 K, and for $V_G$ set at a MRF minimum, there is a increase in $R(T)$. This is strong evidence in favour of our proposal that the varying sign of the resistance temperature dependence below 70 K is due to the presence of MRFs at low $T$ (of either sign) that disappear at high temperatures.

We model our $R(T)$ data for the high charge carrier density region by using the earlier expressions[1,2] to describe our data at higher temperatures, and adding another term to account for the effect of MRFs. Since the rms magnitude of our resistance fluctuations over a range of gate voltages decreases exponentially with temperature, we expect that the individual fluctuation features at a fixed gate voltage should follow the same exponential decay and write for the high charge density regions:

$$R = R_f \exp(-T/T_f) + R_0 + aT + p/(\exp(E_p/k_B T) - 1). \tag{1}$$

Here $R_f$ is the amplitude of the resistance fluctuation in the low-$T$ limit and $T_f$ determines the scale of the decay with temperature. Also included are the residual resistance $R_0$, independent of temperature[1,2], the term $aT$ linear in temperature due to scattering by acoustic phonons[20] and a Bose-Einstein function representing scattering by the dominant high energy phonons[2]. Equation (1) gives a very good account of our data. Replacing the Bose-Einstein function by a $bT^5$-term[1] also gave a very good fit as shown in Figure 3.

Regarding our key result in this Letter, the fits confirm that the exponential decay term representing the decrease of the MRF amplitude (due to loss of phase coherence with $T$) reflects the behaviour in all our experimental data very well. The mean value of the temperature decay constant $T_f$ for these fits is 20 K.

Turning briefly to the resistance at higher temperatures, the fits we show in Figure 3 for the Bose-Einstein function are for longitudinal zone-boundary phonons of energy $E_p$ = 160 meV, since these phonons couple strongly to the electrons, and we have already shown that these 160 meV phonons account well for the somewhat similar increase of resistivity of metallic single-wall carbon nanotube (SWCNT) networks near room temperature[21]. In addition, these 160 meV phonons have been found[22] responsible for the increase in resistance $R(V)$ at high voltages, of individual SWCNTs. However, using the double Bose-Einstein function of Chen et al.[2] for the $SiO_2$ phonons gives a fit similar but no better, so we cannot confirm the energy of the phonons involved.

For the low carrier density region near the NP, our scenario of exponential-like decays of MRFs (positive and negative) also accounts well for the our low $T$ data (Figures 1 and 2). The key difference compared to our data for high charge carrier densities is that the resistance near the NP *decreases* sharply with temperature above 120 K (the conductance shows an exponential-like increase, suggesting some kind of activation effect). We can model the resistance by the same MRF and linear term as at high charge densities, combined with the conductance activation term $\exp(-E_c/k_B T)$ which obviously dominates over the effect of scattering by high-energy phonons (the latter term is omitted):

$$R = [R_f \exp(-T/T_f) + R_0 + aT]/[1 + c(\exp(-E_c/k_B T))]. \qquad (2)$$

Fits to this equation are shown in Figure 2 with fitted values of activation energies $E_c \sim 50$ meV.

In conclusion, we have demonstrated that the sign of the low-$T$ increase or decrease of resistance in graphene depends on whether scattered wave interference is constructive or destructive for the particular pattern of electron scatterers. The fact that the relatively large low-$T$ anomaly changes sign for the same sample at different values of gate voltage is strongly supportive of our identification of scattered wave function interference as the source of the anomalous low-temperature behaviour. Together, Equations (1) ad (2) with the MRF exponential decay term provide a key component for understanding conduction at low temperature for all charge carrier densities.

**METHODS**

The graphene sample was prepared by depositing graphene flakes from a crystal of highly oriented pyrolytic graphite (HOPG) on top of the 300 nm thick $SiO_2$-layer formed on the surface of a heavily doped Si substrate. With the help of a marker coordinate system, the thickness of each graphene flake was then identified by an optical microscope[23,24]. From the optical contrast of digitized images, the number of layers for individual graphene samples was estimated. A monolayer strip of 10 μm in length and ~1.45 μm in width was chosen for our electrical measurements (Fig. 1). A set of seven parallel Cr/Au (3nm/40nm) electrodes was fabricated by the electron beam lithography technique.

Electrical measurements were performed in the 4-probe configuration where electrodes 1 and 7 (see Supplementary information in Fig. S1) were connected to a source-meter Keithley 2400 of constant current (100 nA was used in all measurements) and voltage drops between four inner pairs of electrodes were recorded by four voltmeters Keithley 2000. For the purpose of this study, only the data obtained from the electrode pair 5-6 with the largest distance of 1200 nm will be presented. The bulk silicon substrate, electrically separated from the graphene sample by the $SiO_2$-layer, was connected to a voltage source Keithley 2400 and was used as a back gate for the field effect measurements. The sample was placed into the sample chamber and annealed at 120 $^{o}$C in vacuum ($10^{-7}$mbar) for two days. Then helium gas was introduced into the sample chamber from the helium line, thus ensuring that the pressure of helium gas remain relatively stable in the whole temperature range.

Acknowledgements

V.S. and S.R. acknowledge support from the EC project SANES and to the Project SALVE by Deutsche Forschungsgemeinschaft. V.S. acknowledge the Center of Excellence CENAMOST (Slovak Research and Development Agency Contract No. VVCE-0049–07) with support of project 06–628.

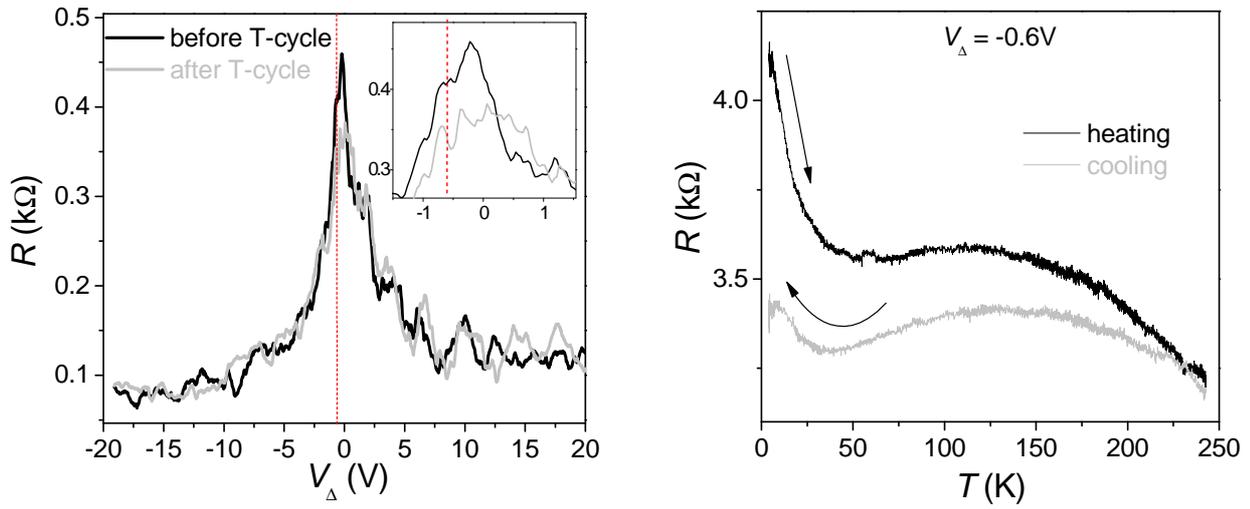

**Figure 1 Resistance of graphene as a function of gate voltage and temperature.**
**a,** Resistance of the graphene flake as a function of gate voltage at 4.2 K before (black) and after (grey) a T-cycle up to 245 K and back down again to 4.2 K.  **b,** Resistance during the T-cycle carried at a constant gate voltage $V_\Delta = -0.6$ V (indicated by dashed lines in **a**).

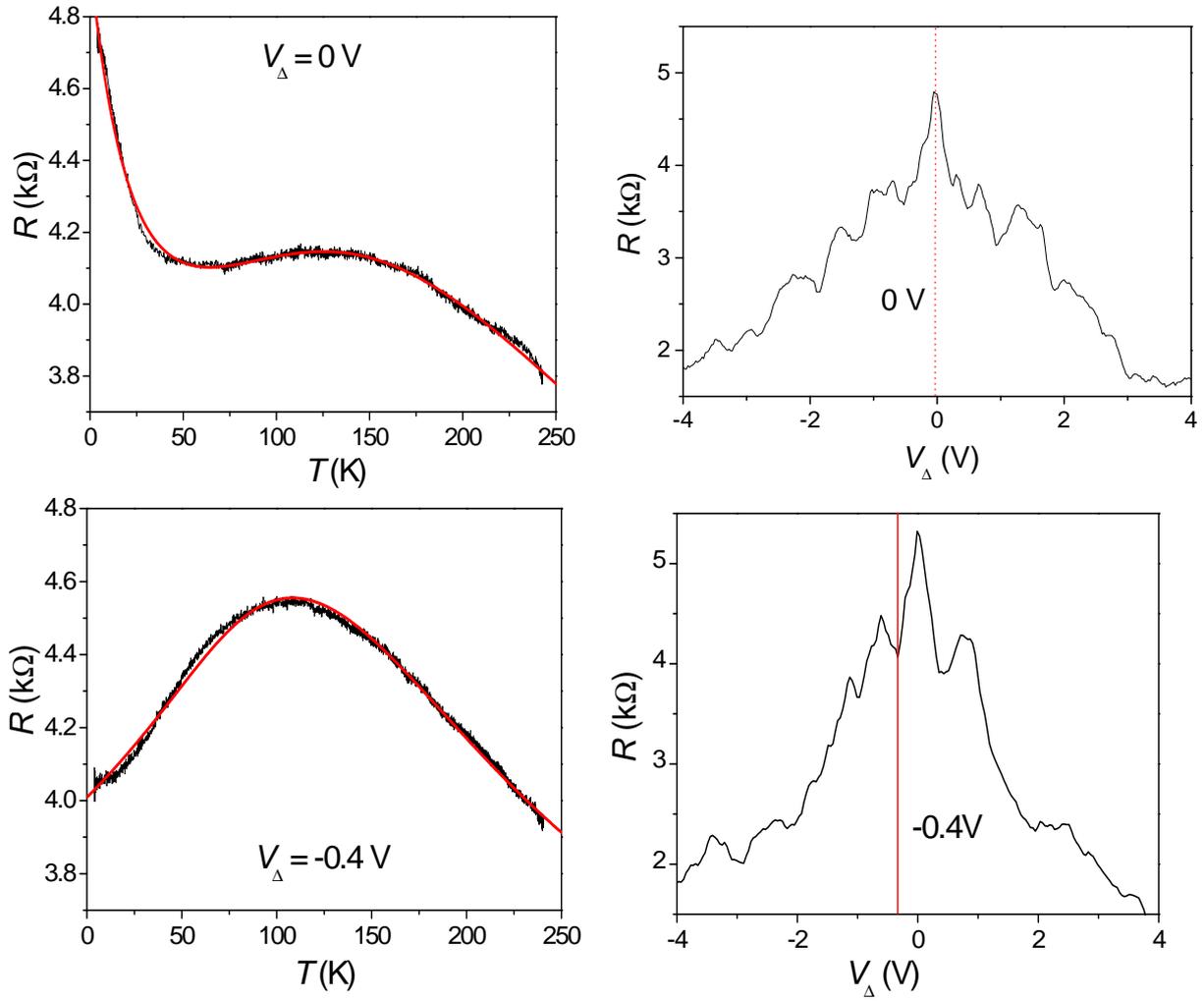

**Figure 2 Resistance of graphene close to the neutrality point.**
**a,** Resistance as a function of temperature at gate voltage $V_\Delta = 0$ at the MRF maximum value at the NP. **b,** Indication of the MRF maximum at $V_\Delta = 0$ by the red line in the resistance dependence on gate voltage.
**c,d,** Same plots as for **a,b,** at a nearby resistance fluctuation minimum value at $V_\Delta = -0.4$ V. The fit lines in **a** and **c** are fits of the $R(T)$ data to the model of Eq. (2), as discussed in the text following equation (2).

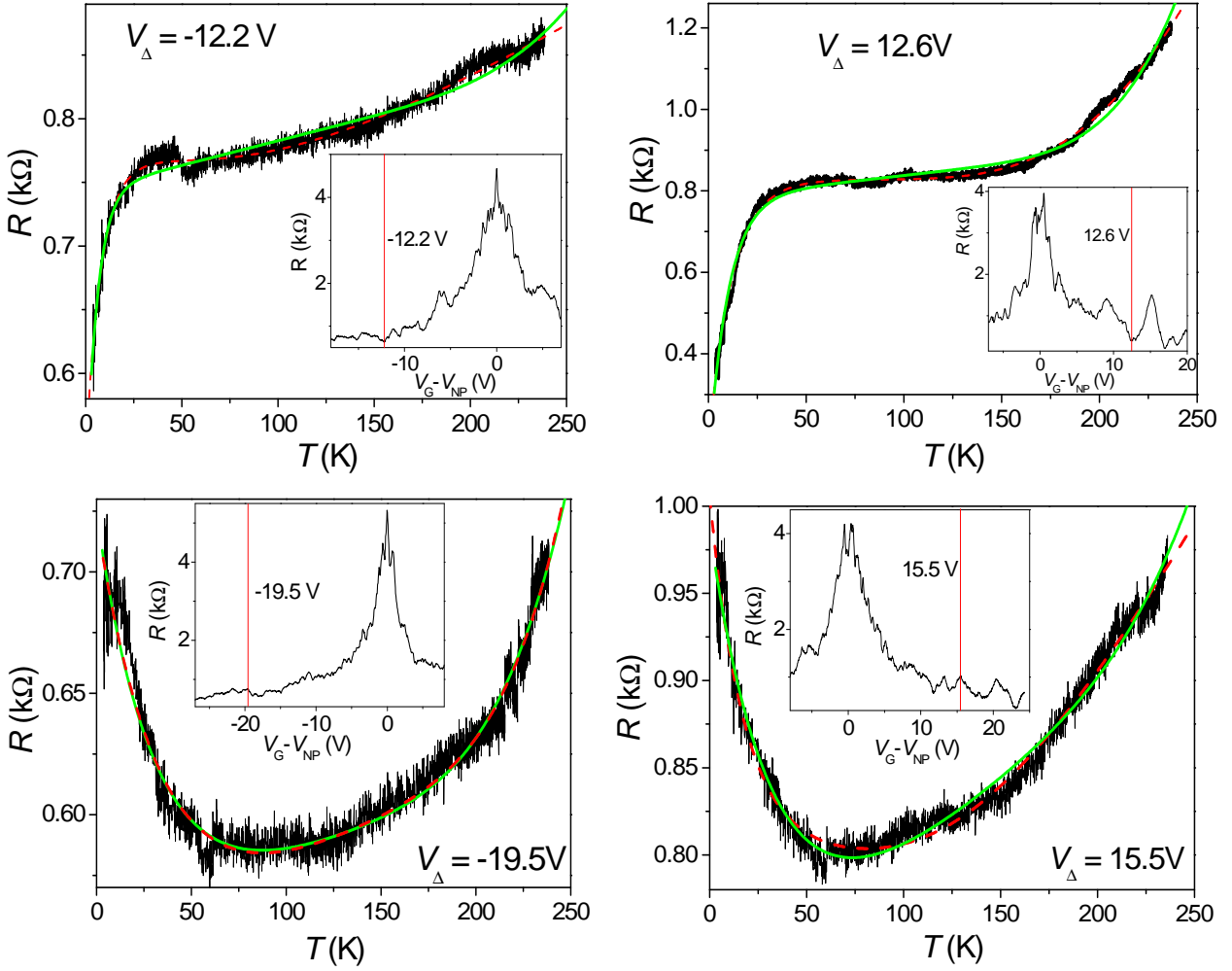

**Figure 3 Resistance of graphene far from the neutrality point**
**a,b,** Resistance as a function of temperature at two different values of the gate voltage far from the NP (i.e. at large charge carrier densities) at minima of the resistance fluctuation pattern (as indicated in the insets). **c,d,** Same plots as **a,b,** but at maxima of the resistance fluctuation pattern. The green solid lines in each panel are fits of the $R(T)$ data to equation (1) where $E_p = 160$ meV for in-plane zone-boundary phonons; the dashed red lines are fits using a $T^5$ term instead of the Bose-Einstein term in equation (1) as discussed in the text.

**Supplementary Information**

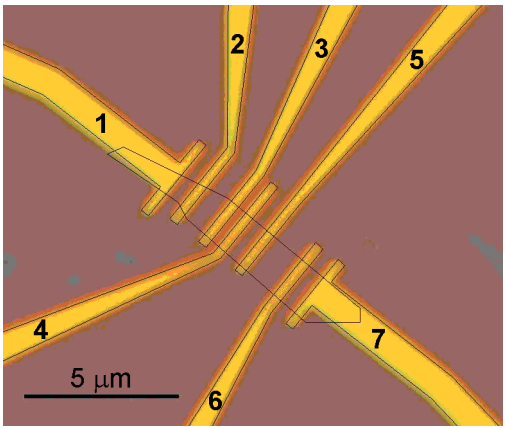

**Figure S1** Optical image of the pattern of gold electrodes on top of a monolayer graphene sample (indicated by the dotted lines).